\documentclass[]{revtex4}

\usepackage{bm}
\usepackage{amsmath,amsfonts}
\begin{document}

\title{Mechanics of  electromagnetic interactions}
\author{Valery P. Dmitriyev}
\affiliation{Lomonosov University\\
P.O.Box 160, Moscow 117574, Russia}
\email{aether@yandex.ru}
\date{\today}

\begin{abstract}
We consider an elastic-plastic medium whose motion equations are isomorphic to Maxwell's equations. Electrical charges are modeled by pressure centers of the medium. The electric interaction is shown to be concerned with the conservation law in the torsion field of the medium. The Lorentz force may correspond to the Coriolis driving force due to the entrainment of the pressure center in the medium's flow.
\end{abstract}
%\pacs{46.05.+b General theory of continuum mechanics of solids, 62.20.F- Deformation and plasticity, 03.50.De Classical electromagnetism,
%Maxwell equations, 12.60.-i Models beyond the standard model, 45.20.Jj Lagrangian mechanics, 47.27.em  Reynolds stress modeling}
\maketitle

\section{Substratum for physics}

The hypothetical substratum underlying classical electromagnetism was shown \cite{Dmitr} to be an elastic-plastic jelly-like medium. The peculiar feature of the jelly is that it is stiff to compression yet liable to shear deformations. Its motion is governed by the following linear equations. The continuity equation for the incompressible medium is
\begin{equation}
{\bm\nabla}\cdot{\bf u} = 0 \label{continuity}
\end{equation}
where ${\bf u}({\bf x},t)$ is the velocity field of the medium. Dynamics of the jelly is described in the Lagrange representation by the type of the Lame equation
\begin{equation}
\varsigma\partial_t{\bf u}+\varsigma c^2{\bm\nabla}\times({\bm\nabla}\times{\bf s})+{\bm\nabla}p-\varsigma{\bf f}=0\label{dynamics}
\end{equation}
where ${\bf s}({\bf x},t)$ is the small displacement of the element of the medium,
\begin{equation}
{\bf u}=\partial_t{\bf s},\label{velocity}
\end{equation}
 $\varsigma$ the density of the medium, $c$ the rate of the transverse wave spreading in the medium and ${\bf f}$ is an external force. Expression (\ref{dynamics}) differs from the standard Lame equation by that for the incompressible medium we have the pressure $p$ term instead of ${\bm\nabla}\cdot{\bf s}$. In the medium under consideration the pressure gradient, i.e. the longitudinal stress, is created by a stationary inclusion that is described by the static external force. While the transverse stress given by the solenoidal term $c^2{\bm\nabla}\times({\bm\nabla}\times{\bf s})$ is of a transitory character. In order to describe this feature we must reformulate (\ref{dynamics}) into the kind of the Prandtl-Reuss model of the plastic-elastic medium which specifies the joint evolution of the velocity ${\bf u}({\bf x},t)$ and external force $-{\bf R}({\bf x},t)$ fields:
\begin{eqnarray}
\varsigma\partial_t{\bf u}+{\bm\nabla}p+\varsigma{\bf R}&=&0,\label{dynamics u}\\
\partial_t{\bf R}-c^2{\bm\nabla}\times({\bm\nabla}\times{\bf u})+{\bf J}&=&0\label{dynamics R}
\end{eqnarray}
 where ${\bf J}({\bf x},t)$ is a current for the external force. By (\ref{dynamics R}) ${\bf u}$ is a small quantity though according to (\ref{velocity}) the displacement ${\bf s}$ may be large. This expresses the material flow in the elastic-plastic medium. The medium in question is realized in the ideal turbulent fluid with the elasticity $c^2{\bm\nabla}\times({\bm\nabla}\times{\bf u})$ and volume force ${\bf R}$ arising due to Reynolds stresses \cite{Dmitr1}.
 Now we have three vector equations, (\ref{continuity}), (\ref{dynamics u}) and (\ref{dynamics R}), for three vector and one scalar unknown fields, ${\bf u}$, ${\bf R}$, ${\bf J}$ and $p$. The set of equations  (\ref{continuity}), (\ref{dynamics u}) and (\ref{dynamics R}) will be closed if we know the dependence of ${\bf J}$ on ${\bf u}$ and $p$. In the case of the electromagnetic substratum such a closure can be performed in the following way.

 The inhomogeneous field ${\bf R}({\bf x},t)$ is generated in the medium by the point defect that is described by the singularity
\begin{equation}
{\bm\nabla}\cdot{\bf R}=4\pi b \delta({\bf x}-{\bm\xi})\label{center}
\end{equation}
where $b$ is the strength of the defect and ${\bm\xi}$ its location  \cite{Dmitr}. We suppose that the discontinuity (\ref{center}) can move freely in the medium its motion velocity being defined as
\begin{equation}
{\bf v}=\frac{d{\bm\xi}}{dt}.\label{center velocity}
\end{equation}
The motion of defects underlies the phenomenology of the plasticity. A moving with the velocity (\ref{center velocity}) singularity (\ref{center}) forms in the medium the current of the defect's strength \cite{Dmitr}:
\begin{equation}
{\bf J}= 4\pi b{\bf v}\delta({\bf x}-{\bm\xi}).\label{current}
\end{equation}
We assign to the defect (\ref{center}) the mass $m$ and assume that it moves in the medium as a classical particle:
\begin{equation}
m\frac{d{\bf v}}{dt}={\bf F}.\label{particle dynamics}
\end{equation}
Now the problem of the closure for the set of equations (\ref{continuity}), (\ref{dynamics u})-(\ref{particle dynamics}) will be to find the dependence of ${\bf F}$ on $b$ and the fields ${\bf u}$ and $p$, and also on ${\bf v}$. In physical terms we are searching for the law that defects interact with each other by means of the fields that they generate in the medium.

\section{Interaction due to conservation law in the torsion field}

Take notice that multiplying the current (\ref{current}) by any vector field ${\bf R}({\bf x},t)$ and integrating all over the volume we will obtain the expression
\begin{equation}
\int{\bf J}\cdot{\bf R}d^3x=4\pi b{\bf v}\cdot{\bf R}({\bm\xi},t)\label{work}
\end{equation}
that depends only on coordinates of the defect (\ref{center}) referred here as the particle. The form (\ref{work}) can be interpreted as the rate of the work done on the particle
\begin{equation}
\frac{\kappa^2}{4\pi}\int{\bf J}\cdot{\bf R}d^3x={\bf v}\cdot{\bf F}({\bm\xi},t)\label{power work}
\end{equation}
if we will assume that the force acted on the particle is
\begin{equation}
{\bf F}=\kappa^2b{\bf R}\label{force}
\end{equation}
where $\kappa^2$ is a positive constant.

Basing himself on this observation we will multiply (\ref{dynamics R}) by the term ${\bf R}$ of the external field and integrate it over the volume:
\begin{equation}
\frac{1}{2}\partial_t\int{\bf R}^2d^3x-c^2\int{\bf R}\cdot[{\bm\nabla}\times({\bm\nabla}\times{\bf u})]d^3x+\int{\bf J}\cdot{\bf R}d^3x=0.\label{dynamics R by R}
\end{equation}
Using (\ref{dynamics u}) in the second term of (\ref{dynamics R by R}) and integrating it by parts gives
\begin{equation}
-\int{\bf R}\cdot[{\bm\nabla}\times({\bm\nabla}\times{\bf u})]d^3x=\int(\partial_t{\bf u}+{\bm\nabla}p)\cdot[{\bm\nabla}\times({\bm\nabla}\times{\bf u})]d^3x=\frac{1}{2}\partial_t\int({\bm\nabla}\times{\bf u})^2d^3x\label{second term}
\end{equation}
where vector relation ${\bf B}\cdot({\bm\nabla}\times{\bf A})={\bf A}\cdot({\bm\nabla}\times{\bf B})+{\bm\nabla}\cdot({\bf A}\times{\bf B})$ and $\partial_t{\bf u}\times({\bm\nabla}\times{\bf u})\rightarrow0$ at infinity were used. Substituting (\ref{power work}) and (\ref{second term}) into (\ref{dynamics R by R}):
\begin{equation}
\frac{\kappa^2}{8\pi}\partial_t\int[{\bf R}^2+c^2({\bm\nabla}\times{\bf u})^2]d^3x+{\bf F}\cdot{\bf v}=0.\label{dynamics R by R+}
\end{equation}
Using (\ref{particle dynamics}) in (\ref{dynamics R by R+}) gives
\begin{equation}
\frac{d}{dt}\left\{\frac{\kappa^2}{8\pi}\int[{\bf R}^2+c^2({\bm\nabla}\times{\bf u})^2]d^3x+\frac{mv^2}{2}\right\}=0.\label{conservation}
\end{equation}
So, with a reasonable assumption (\ref{force}) for the force on the particle we have balance (\ref{conservation}) of the particle's kinetic energy within the integral in the torsion field of the medium. The terms with ${\bf R}^2$ and $c^2({\bm\nabla}\times{\bf u})^2$ in (\ref{conservation}) can be interpreted as the energy of respective fields concerned with the energy of the particle. However, as we see, they have no relation to the energy of the substratum which is proportional to ${\bf u}^2$ and an integral of ${\bf R}$ over a space coordinate. Using (\ref{dynamics u}) in (\ref{force}) gives for the field force
\begin{equation}
{\bf F}=-\kappa^2b(\partial_t{\bf u}+{\bm\nabla}p/\varsigma).\label{field force}
\end{equation}

\section{Driving force due to entrainment of the particle in the flow of the medium}

We will assume that the singularity (\ref{center}) is entrained in the flow of the medium. So that the full velocity of the defect can be represented as the sum of the defect's own velocity and the medium's velocity. In a more general case the defect may be a composite structure. We will assume that it moves as a classical particle of the mass $m$ with $\mu$ being its entrained part, $m\geq\mu$. Then the full velocity of the particle  ${\bf v}$ can be represented as the sum of the particle's own velocity ${\bm\upsilon}$ and the part of the medium's velocity
\begin{equation}
{\bf v}={\bm\upsilon}+\frac{\mu}{m}{\bf u}\label{entrained}
\end{equation}
where $\mu$ is the mass coefficient of the pressure center (\ref{center}). Suppose that the kinematics of the field ${\bf u}({\bf x},t)$ is known. Then we may thus or otherwise compute  the driving force on the particle. In the absence of a specific force it is formally given by
\begin{equation}
{\bf F}=m\frac{d{\bf v}}{dt}.\label{driving}
\end{equation}
As can be shown (see Appendix) the driving force on the particle consists of the inertial force plus the force due to rotation of the particle's velocity by the medium's flow:
\begin{equation}
{\bf F}={\mu}\frac{d{\bf u}}{dt}-{\mu}{\bm\upsilon}\times({\bm\nabla}\times{\bf u}).\label{driving force}
\end{equation}

In order to interpret terms in the right-hand part of (\ref{driving force}) we will apply (\ref{driving force}) to the well-known particular case, the rotational field
\begin{equation}
{\bf u}=\frac{d{\bf x}}{dt}={\bm\omega}\times{\bf x},\label{rotation}
\end{equation}
We have for (\ref{driving force}) with (\ref{rotation})  in two dimensions
\begin{eqnarray}
\frac{d{\bf u}}{dt} &=& -\omega ^2 {\bf x},\label{centripetal}\\
{\bm\nabla}\times{\bf u} &=& 2{\bm \omega }.\label{Coriolis}
\end{eqnarray}
Substituting (\ref{Coriolis}) into the last term of (\ref{driving force}) we obtain the expression for the Coriolis driving force on the particle.
Relation (\ref{centripetal}) used in (\ref{driving force}) gives the centripetal driving force on the particle.

In the linear-elastic medium the full time derivative in (\ref{driving force}) can be changed to partial derivative and the full particle's velocity ${\bf v}$ substituted for its relative velocity ${\bm\upsilon}$:
\begin{equation}
{\bf F}={\mu}\partial_t{\bf u}-{\mu}{\bf v}\times({\bm\nabla}\times{\bf u}). \label{driving force Lagrange}
\end{equation}
With some suggestions concerning the mass $\mu$ of the pressure center (\ref{center}) relation (\ref{driving force Lagrange}) provides an addition to the specific force in (\ref{field force}) found from field equations.

\section{Entrainment in various systems}

We will consider three variants for the entrainment of a particle in a medium's flow.

Let an impurity be advected by the laminar flow of an ideal fluid according to (\ref{entrained}). Using in (\ref{driving force}) the Euler equation
\begin{equation}
\varsigma\frac{d{\bf u}}{dt}+{\bm\nabla}p=0\label{Euler}
\end{equation}
 we obtain
\begin{equation}
{\bf F} = -\frac{\mu}{\varsigma}{\bm\nabla}p-\mu{\bm\upsilon}\times({\bm\nabla}\times{\bf u}). \label{Euler driving force}
\end{equation}
Relations (\ref{centripetal})-(\ref{Coriolis}) provide a particular case to (\ref{Euler driving force}).

Next we consider transport in the elastic-plastic medium. Using (\ref{dynamics u}) in (\ref{driving force Lagrange}) we obtain
\begin{equation}
{\bf F} = -\mu(\frac{1}{\varsigma}{\bm\nabla}p+{\bf R})  -
\mu{\bf v}\times({\bm\nabla}\times{\bf u}). \label{driving force elastic}
\end{equation}
We see that there may be a regime when the external force fully equilibrates the pressure in (\ref{driving force elastic}). The result will be a purely rotation of the particle's velocity.

We have a special case in the electromagnetic substratum. Combining (\ref{driving force Lagrange}) with (\ref{field force}) gives with the account of (\ref{dynamics u}) for the full force on the singularity
\begin{equation}
{\bf F}=\kappa^2b{\bf R}+\kappa^2b{\bf v}\times({\bm\nabla}\times{\bf u})\label{full force}
\end{equation}
where the mass coefficient $\mu$ in (\ref{driving force Lagrange}) is related with the strength $b$ of the pressure center (\ref{center}) by
\begin{equation}
\mu = -\kappa^2b.\label{mass}
\end{equation}
As in contrast with (\ref{Euler driving force}) and (\ref{Euler}) we may have in (\ref{full force}) and (\ref{dynamics u}) separately the Coriolis force and the central force. This situation becomes possible due to the excluding in (\ref{Euler}) of non-linear terms when passing to Lagrange coordinates. The addition of the Coriolis driving force to ({\ref{field force}) does not change the result of the last term in (\ref{dynamics R by R+}).

\section{Electromagnetic analogy}

We define vector and scalar electromagnetic potentials
\begin{eqnarray}
{\bf A} &=& \kappa c{\bf u}\label{vector potential},\\
\varsigma\varphi&=&\kappa(p-p_0),\label{scalar potential}\\
{\bf E}&=&\kappa{\bf R}\label{electric}
\end{eqnarray}
where $p_0$ is the background pressure. Substituting (\ref{vector potential})-(\ref{electric}) into (\ref{dynamics u}) we will obtain
the Maxwell's equation
\begin{equation}
\frac{1}{c}\partial_t{\bf A}+{\bf E}+{\bm\nabla}\varphi=0.\label{Maxwell}
\end{equation}
With the substitution (\ref{vector potential}) continuity equation (\ref{continuity}) takes the form of the Coulomb gauge
\begin{equation}
{\bm\nabla}\cdot{\bf A} = 0. \label{gauge}
\end{equation}
 We will define the electric charge by
\begin{equation}
q = \kappa b.\label{charge}
\end{equation}
Substituting (\ref{electric}) and (\ref{charge}) into (\ref{center}) we obtain the source equation
\begin{equation}
{\bm\nabla}\cdot{\bf E}=4\pi q \delta({\bf x}-{\bm\xi}).\label{source}
\end{equation}
Substituting (\ref{vector potential}), (\ref{electric}) and (\ref{charge}) into (\ref{dynamics R}) with (\ref{current}) gives another Maxwell's equation
\begin{equation}
\partial_t{\bf E}-c{\bm\nabla}\times({\bm\nabla}\times{\bf A})+ 4\pi q{\bf v}\delta({\bf x}-{\bm\xi}) = 0.\label{Maxwell field evolution}
\end{equation}
Substituting (\ref{electric}), (\ref{vector potential}) and (\ref{charge}) into (\ref{full force}) we come to the expression for the force acted on the electric charge
\begin{equation}
{\bf F} = q[{\bf E}+\frac{{\bf v}}{c}\times({\bm\nabla}\times{\bf A})]. \label{electromagnetic force}
\end{equation}

The proton $q>0$ is modeled mesoscopically by a cavity stabilized via the perturbation of the field ${\bf R}$ \cite{Dmitr1}. Through (\ref{center}) and (\ref{dynamics u}) with (\ref{continuity}), the decreased pressure in the center of the defect corresponds to $b<0$ . Hence by (\ref{charge}) $\kappa<0$. Comparing (\ref{charge}) with (\ref{mass}) we may conclude that
\begin{equation}
\kappa=-\frac{\mu}{e}\label{coefficient}
\end{equation}
where $\mu$ is the mass coefficient and $e$ the charge of the electron.

By (\ref{coefficient}) with $\kappa<0$ we have $\mu<0$ for $e<0$. This means that the electron will move counter to the aether flow. The loop on a vortex filament microscopic mechanism that underlies the upstream motion of the particle was considered in my earlier publication \cite{Dmitr2}.

\section{Realization}

The medium that provides substratum for physics can be realized in the ideal turbulent fluid \cite{Dmitr1}. Following Osborne Reynolds we expand the fluid velocity into the sum of the averaged $\left<{\bf u}\right>$ over a short time interval and fluctuation ${\bf u}'$ components:
\begin{equation}
{\bf u}=\left<{\bf u}\right>+{\bf u}'\label{turbulence}
\end{equation}
where
\begin{equation}
\left<{\bf u}'\right>=0.\label{fluctuation}
\end{equation}
Substituting (\ref{turbulence}) into (\ref{Euler}), averaging with the account of (\ref{fluctuation}) and linearizing gives \cite{Troshkin}
\begin{equation}
\varsigma\partial_t\left<{\bf u}\right>+\varsigma\partial_j\left<{\bf u}'u'_j\right>+{\bm\nabla}\left<p\right> = 0\label{linear Reynolds}
\end{equation}
where $\partial_iu_i'=0$ was used. The summation over recurrent index is implied here and throughout. Comparing (\ref{linear Reynolds}) with (\ref{dynamics u}) we see that all the fields mentioned can be related to averaged perturbations of the ideal turbulence, and the external force $\bf R$ can be rendered in terms of Reynolds stresses:
\begin{equation}
{\bf R} = \partial_j\left<{\bf u}'u'_j\right>.\label{turbulence force}
\end{equation}
 Using (\ref{turbulence force}) in (\ref{electric}) we find that the electric field is concerned with the turbulence force acted on the element of the medium:
\begin{equation}
{\bf E}=\kappa\partial_j\left<{\bf u}'u'_j\right>\label{electric field turbo}
\end{equation}
where with the account of (\ref{coefficient}) $\kappa<0$.  The meaning of (\ref{electric field turbo}) is that the  electromotive force is proportional to the difference of the turbulence energy density $\left<u'_iu'_i\right>/2$ across two open-circuited terminals. Alike in (\ref{electric field turbo}), definitions of vector (\ref{vector potential}) and scalar (\ref{scalar potential}) potentials should be changed to respective averaged values.
Using (\ref{turbulence force}) in the first term of (\ref{full force}) we obtain for the specific force acted on the particle
 \begin{equation}
F_i=-|\kappa|q\partial_j\left<u'_iu'_j\right>\label{turbophoresis}
\end{equation}
where (\ref{charge}) and $\kappa<0$ were used. As was above said, $q>0$ corresponds to a cavity stabilized in the disturbed turbulence. According to (\ref{turbophoresis})  we may expect that the cavity will move in the region of the decreased turbulence intensity. This tendency just corresponds to the phenomenon in macroscopic transport known as turbophoresis. When $q>0$ the dependence (\ref{turbophoresis}) coincides with the expression for the force on a large particle derived theoretically in \cite{Reeks}. The largeness of the pressure center (\ref{center}) consists in that it generates in the medium the voluminous perturbation field $-b/r$. The inversion of the turbophoretic force with the change of the sign of the pressure center suggests that the turbophoresis may have the same vortex loop microscopic mechanism as the Lorentz force \cite{Dmitr2}

In order to reproduce properly the Lorenz force in (\ref{electromagnetic force}) we must assume in (\ref{entrained}) that the pressure center is entrained by the averaged flow, i.e.
\begin{equation}
{\bf v} = {\bm\upsilon} +  \frac{\mu}{m}\left<{\bf u}\right>,\label{full velocity averaged}
\end{equation}
and then average the force (\ref{driving force Lagrange}) acted on the particle.

%\enlargethispage{10pt}

\appendix
\section{The entrainment force}
Let the particle at ${\bm\xi}$ is driven by the flow of a medium. The kinematics of the medium is specified by the velocity field ${\bf u}({\bf x},t)$. We will assume that at any instant of time the full velocity ${\bf v}$ of the particle can be represented as the sum of the particle's own velocity (relative velocity) ${\bm\upsilon}$ and the medium's velocity:
\begin{equation}
{\bf v} = {\bm \upsilon} +  {\bf u}({\bm \xi},t).\label{full velocity A}
\end{equation}
The kinematics of the particle can be calculated if we know the dependence on time of the particle's relative velocity in the reference frame
${\bf {\check x}}$ accompanying the medium's velocity field at ${\bm\xi}$:
\begin{equation}
{\bm{\check\upsilon}} = \frac{d{\bm{\check\xi}}}{dt}.\label{relative velocity}
\end{equation}
We will assume that the relative motion of the particle is governed by a specific force ${\bm \Re}$:
\begin{eqnarray}
\frac{d{\bm{\check\upsilon}}}{dt} = {\bm{\check\Re}}.\label{non-entrainment}
\end{eqnarray}

Considering Lagrange dynamics in the non-inertial reference frame ${\bf\check x}$ we will compute the driving force ${\bf F}-{\bm\Re}$ on the particle due to the entrainment (\ref{full velocity A}). The kinetic energy of the particle in the reference frame ${\bf\check x}$ is
\begin{equation}
{\check K} = \frac{1}{2}m[{\bm{\check\upsilon}}(t) +  {\bf{\check
u}}({\bm{\check\xi}},t)]^2\label{kinetic}
\end{equation}
where
\begin{equation}
 {\bf{\check u}}({\bm{\check\xi}}(t),t)={\bf u}({\bm{\check\xi}}(t),t)-{\bf u}({\bf{\check x}}).\label{new velocity}
\end{equation}
Here we do not indicate explicitly the time dependence in the velocity field ${\bf u}({\bf x})$ of the medium since this term will not be differentiated with respect to time. Lagrange equations for the evolution of the particle's relative velocity (\ref{relative velocity}) are
\begin{equation}
\frac{d}{{dt}}\frac{{\partial\check K}}{\partial{{\bm{\check\upsilon}}}} = \frac{{\partial\check K}}{\partial{\bm{\check\xi}}}
+ {\bf{\check F}}.\label{Lagrange}
\end{equation}
Substituting  (\ref{kinetic}) in (\ref{Lagrange}) we have with the account of (\ref{relative velocity})
\begin{equation}
\frac{d}{{dt}}\frac{{\partial\check K}}{\partial{\bm{\check\upsilon}}}
= m\frac{d}{{dt}}\left({\bm{\check\upsilon}} + {\bf{\check u}}
\right) = m\left[\frac{d{\bm{\check\upsilon}}}{dt} + \left(
{{\bm{\check\upsilon}} \cdot {\bm{\check\nabla}}}\right){\bf{\check u}}+\frac{d{\bf{\check u}}}{dt}\right],\label{left Lagrange}
\end{equation}
where
\begin{equation}
\frac{d{\bf{\check u}}}{dt} = \frac{\partial{\bf{\check u}}}{\partial t}+\left(
{{\bf{\check u}} \cdot {\bm{\check\nabla}}}\right){\bf{\check u}},
\label{derivative}
\end{equation}
and
\begin{equation}
\frac{{\partial\check K}}{\partial{\bm{\check\xi}}} =
\frac{1}{2}m{\bm{\check\nabla}}\left({\bm{\check\upsilon}} +
{\bf{\check u}} \right)^2 = m\left({{\bm{\check\upsilon}}\cdot
{\bm{\check\nabla}}} \right){\bf{\check u}} + m{\bm{\check\upsilon}} \times ({\bm{\check\nabla}}\times{\bf{\check u}}) +
\frac{1}{2}m{\bm{\check\nabla}}{\bf{\check{u}}}^2,\label{right Lagrange}
\end{equation}
where the vector relation ${\bm\nabla} \left( {{\bf a}
\cdot {\bf b}} \right) = \left( {\bf a} \cdot {\bm\nabla}
\right){\bf b} + \left( {{\bf b} \cdot {\bm\nabla} } \right){\bf
a} + {\bf a} \times ({\bm\nabla}\times{\bf b}) + {\bf b} \times
({\bm\nabla}\times{\bf a})$ was used. Substituting  (\ref{left Lagrange}) and
(\ref{right Lagrange}) into (\ref{Lagrange}) we obtain
\begin{equation}
m\frac{d{\bm{\check\upsilon}}}{dt}+m\frac{d{\bf{\check u}}}{dt} =
m{\bm{\check\upsilon}}\times({\bm{\check\nabla}}\times{\bf{\check
u}}) + \frac{1}{2}m{\bm{\check\nabla}}{\bf{\check u}}^2 + {\bf{\check F}}.
\label{new equation}
\end{equation}
Using (\ref{non-entrainment}) in (\ref{new equation}) gives for the driving force of the entrainment
\begin{equation}
{\bf{\check F}}-{\bm{\check\Re}}=m\frac{d{\bf{\check u}}}{dt} - \frac{1}{2}m{\bm{\check\nabla}}{{\bf{\check u}}}^2 -
m{\bm{\check\upsilon}}\times({\bm{\check\nabla}}\times{\bf{\check
u}}). \label{new driving force}
\end{equation}
Next we may convert (\ref{new driving force}) into the inertial reference frame ${\bm{\check\xi}}\rightarrow{\bm\xi}$  and ${\bf{\check x}}\rightarrow{\bf x}$ taking advantage of that ${\bm{\check\nabla}}\rightarrow{\bm\nabla}$, ${\bf{\check F}}\rightarrow{\bf F}$ and ${\bm{\check\upsilon}}\rightarrow{\bm\upsilon}$. Substituting (\ref{new velocity}) into (\ref{new driving force}) gives
\begin{equation}
{\bf F} = {\bm\Re}+m\frac{d{\bf u}({\bm\xi}(t),t)}{dt} - \frac{1}{2}m{\bm\nabla}[{\bf u}({\bm\xi}(t),t)-{\bf u}({\bf x})]^2 -
m{\bm\upsilon}\times[{\bm\nabla}\times{\bf u}({\bm\xi}(t),t)]. \label{driving force-}
\end{equation}
Taking ${\bf x}={\bm\xi}$ the gradient of the quadratic term in (\ref{driving force-}) will be nullified, and we shall have finally for the force on the particle entrained in the medium
\begin{equation}
{\bf F} = {\bm\Re}+m\frac{d{\bf u}}{dt}  -
m{\bm\upsilon}\times({\bm\nabla}\times{\bf u}). \label{driving force A}
\end{equation}
Specifying dynamical law for ${\bm\Re}$ and ${\bf u}$ in the right-hand part of (\ref{driving force A}) we shall find the force exerted on a particle by a particular medium.

\end{document}